\documentclass[onecolumn,showpacs,preprintnumbers,amsmath,amssymb]{revtex4}
\usepackage{bm,amsmath,amssymb,latexsym,mathrsfs,graphicx,enumerate}
\usepackage[mathcal]{euscript}
\usepackage{hyperref}
\usepackage{epsfig}

\usepackage{amsmath}
\usepackage{graphics}
\begin{document}

\title{\textbf{Electromagnetic Wave Propagation in a Quasi-1D Rhombic Linear Optical Waveguide Array}}
\author{Andrey I. Maimistov$^{1,2}$, Viktor A. Patrikeev$^{1}$}
\affiliation{\normalsize \noindent $^1$: Department of General
Physics, Moscow Institute of Physics
and Technology, Dolgoprudny, Moscow region, 141700 Russia \\
$^2$: Department of Solid State Physics and Nanostructures, National
Nuclear Research University,
Moscow Engineering Physics Institute, Moscow, 115409  \\
E-mails: aimaimistov@gmail.com, sugrobs@yandex.ru \\
}

\date{\today}

\begin{abstract}
\noindent  The quasi-one-dimensional rhombic array of the waveguides
is considered. System of equations describing coupled waves in the
waveguide in the linear limit is solved exactly. The electric field
distribution was found both for the diffractionless (or
dispersionless) flat band modes and for the dispersive modes.

\end{abstract}


\maketitle

%


\section{Introduction}

\noindent Recently the optical simulations of the different
phenomena of solid state have been developed. There is one
interested example. The investigation of two dimensional electron
systems demonstrated that presence of the third atom in the
elementary sell as well as long range interaction in the lattice
leads to emerging of a flat sheet (flat band) between conventional
zones. Similar optical lattices can be realized by means of
waveguides as nodes of the lattice. Some kinds of the optical
lattices that demonstrate the photonic spectrum with flat band have
been discussed in \cite{Longi:14,Maimis:15,Maim:Gabi:16}.

Let us consider wave guide array consisting from three parallel
linear chain of waveguides. The central waveguide chain is shifted
according to either chains at half lattice period. Resulting
configuration seams as linear chain of the rhombus. This array of
waveguides was named as the quasi-one-dimensional rhombic array
\cite{Longi:14,Mukherjee:15,Mukherjee:Spracklen:15a}.

In this article the electromagnetic field distribution in the
quasi-one-dimensional rhombic array of the waveguides is obtained.
All waveguides are supposed as linear waveguides. It allows using
the standard technique of solution of the differential-difference
equations. The exact solution of the coupled mode equations is
found. These solutions demonstrate the discrete diffraction
phenomenon that takes place in general case of the boundary (or
initial) conditions. But in the special case of electromagnetic
field excitation the flat band modes will be existed.

\section{Base equations and dispersion relation of propagating modes}

\noindent Let us consider the quasi-one-dimensional rhombic
waveguide array of waveguides \cite{Longi:14,Mukherjee:15}. The
system of equations describing coupled waves in the array has the
following form
\cite{Mukherjee:15,Mukherjee:Spracklen:15a,Vicencio:Cantillano:15}
\begin{eqnarray}
&& i\left(\frac{\partial}{\partial\tau}+ \frac{\partial}{\partial
\xi} \right)A_n =
(B_{n}+B_{n-1})+(C_{n}+C_{n-1}), \nonumber\\
&& i\left(\frac{\partial}{\partial \tau}+ \frac{\partial}{\partial
\xi} \right)B_{n} = (A_{n+1}+A_{n}),
 \label{eq:romb:ABCn:2} \\
&& i\left(\frac{\partial}{\partial \tau}+ \frac{\partial}{\partial
\xi} \right)C_{n} = (A_{n+1}+A_{n}).
 \nonumber
\end{eqnarray}
Here $A_n $, $B_n $ and $C_n $ are dimensionless slowly varying
amplitudes of the electric fields in the waveguide of the $n$-th
elementary sell, the sub-indices $n$ are markers of the elementary
sells. Phase matching condition is assumed to be satisfied, and the
coupling constants between waveguides are equal to unit. Since
system of equation (\ref{eq:romb:ABCn:2}) is linear, the dispersion
relation can be found in a standard way.

If new variable 
$\zeta$ so that $\partial/\partial\tau +\partial/\partial\xi
=\partial/\partial\zeta$ is introduced, then the system of equation
(\ref{eq:romb:ABCn:2}) can be written as
\begin{eqnarray}
&& i\frac{\partial A_n}{\partial \zeta}  =
(B_{n}+B_{n-1})+(C_{n}+C_{n-1}), \nonumber\\
&& i \frac{\partial B_{n}}{\partial \zeta} = (A_{n+1}+A_{n}),
 \label{eq:romb:ABCn:3} \\
&& i\frac{\partial C_{n}}{\partial \zeta} = (A_{n+1}+A_{n}).
 \nonumber
\end{eqnarray}
Since system of equation (\ref{eq:romb:ABCn:3}) is linear, the
dispersion relation can be found in a standard way:
\begin{equation}\label{eq:romb:dispers:abc:stat}
    q_{\pm}(s) = \pm 2 \sqrt{2}|\cos(\pi s/M)|, \quad q_{0}(s) = 0.
\end{equation}
Here $q(s)$ is transversal wave number of the mode with index $s$,
$s=-M,...,-1,0,1,..., M$ $N=2M+1$ is number of the elementary sells
of waveguide array. As noted in
\cite{Longi:14,Maimis:15,Maim:Gabi:16}, the wave numbers
$q_{\pm}(s)$ are correspond to waves propagating between waveguides
in array. The modes with $q_{0}(s)$ form the flat band.

\section{The solution of the base equations}

\noindent To obtain the solution of the system of equation
(\ref{eq:romb:ABCn:3}) we can use the generation function method.
Let us introduce the following functions
$$ P_A(\zeta,
y)=\sum_{n=-\infty}^{\infty} A_n(\zeta)e^{iyn}, \quad P_B(\zeta,
y)=\sum_{n=-\infty}^{\infty} B_n(\zeta)e^{iyn}, \quad P_C(\zeta,
y)=\sum_{n=-\infty}^{\infty} C_n(\zeta)e^{iyn}.
$$
From (\ref{eq:romb:ABCn:3}) the following system of equations can be
obtained
\begin{equation}\label{eq:romb:ProABCs:sim:1}
    i\frac{\partial }{\partial \zeta}P_A =
\kappa^{*}(P_B +P_C), \quad i\frac{\partial }{\partial \zeta} P_B=
\kappa P_A,\quad i\frac{\partial }{\partial \zeta} P_C= \kappa P_A,
\end{equation}
where
$$
\kappa(y)=\left(1+e^{-iy}\right) =2\cos(y/2)e^{-iy/2}.
$$
These equations can be reduced to one second order differential
equation
$$
\frac{\partial^2 }{\partial \zeta^2}P_A +2 |\kappa|^2 P_A=0.
$$
The general solution of this equation is
$$
P_A(\zeta, y) = C_1(y)e^{i\Omega y}+ C_2(y)e^{-i\Omega y}.
$$
where $\Omega^2 =8\cos^2(y/2)$. The constants of integration $C_1$
and $C_2$ can be found from the boundary conditions
$$
\zeta =0:\quad P_A(0, y) =P_{A0}, \quad \left.\frac{\partial P_A
}{\partial \zeta}\right|_{\zeta=0}=\kappa^{*}(P_{B0} +P_{C0}),$$
where
$$
P_{A0}=\sum_{n=-\infty}^{\infty} A_n(0)e^{iyn}, \quad
P_{B0}=\sum_{n=-\infty}^{\infty} B_n(0)e^{iyn}, \quad
P_{C0}=\sum_{n=-\infty}^{\infty} C_n(0)e^{iyn}.
$$
Taking into account these conditions the equations for constants of
integration can be written as
\begin{eqnarray}
  &&  C_1 + C_2= P_{A0}, \nonumber \\
  &&  C_1 - C_2= -\frac{\kappa^{*}}{\Omega}(P_{B0} +P_{C0}). \nonumber
\end{eqnarray}
That results in
\begin{equation}\label{eq:romb:Pro:sim:PA}
    P_A(\zeta, y) = P_{A0}\cos\Omega\zeta -i\beta(P_{B0}
    +P_{C0})\sin\Omega\zeta,
\end{equation}
where
$$
\beta =\frac{\kappa^{*}}{\Omega}= \frac{1}{\sqrt{2}}e^{iy/2}.
$$

The second and third equations from (\ref{eq:romb:ProABCs:sim:1})
can be rewritten as following ones
\begin{eqnarray*}
  && i\frac{\partial }{\partial \zeta}(P_B +P_C) = 2\kappa P_A,\\
  && i\frac{\partial }{\partial \zeta}(P_B -P_C) = 0.
\end{eqnarray*}
It follows that the function $S=P_B -P_C$ is constant $S_0=P_{B0}
-P_{C0}$. Either function $R=P_B +P_C$ is obtained from the first
equation of (\ref{eq:romb:ProABCs:sim:1}), which is rewritten as:
$$
i\frac{\partial P_A}{\partial \zeta} = \kappa^{*}R.
$$
From this equation it follows that
$$
R(\zeta) = R_0\cos\Omega\zeta
-i\frac{1}{\beta}P_{A0}\sin\Omega\zeta,  \qquad R_0=P_{B0} +P_{C0}.
$$
Taking into account the definitions $ S=P_B -P_C$ è $R=P_B +P_C$,
the generation functions can be expressed by following equations
\begin{eqnarray}
  P_B(\zeta,y) &=& \frac{1}{2} \left[P_{B0} -P_{C0}+ (P_{B0}+P_{C0})\cos\Omega\zeta
   -\frac{i}{\beta}P_{A0}\sin\Omega\zeta\right],\label{eq:romb:Pro:sim:PB} \\
  P_C(\zeta,y) &=& \frac{1}{2} \left[P_{C0} -P_{B0}+ (P_{B0}+P_{C0})\cos\Omega\zeta
   -\frac{i}{\beta}P_{A0}\sin\Omega\zeta\right],\label{eq:romb:Pro:sim:PC}
\end{eqnarray}

If the boundary conditions are choosing in the form $ A_n(0) = 0$, $
B_n(0) = - C_n(0) = 0$ then $P_{B0}= -P_{C0}$ and $P_{A0}= 0$. From
(\ref{eq:romb:Pro:sim:PA}), (\ref{eq:romb:Pro:sim:PB}) and
(\ref{eq:romb:Pro:sim:PC}) it follows that
$$
P_A(\zeta, y)=0, \quad P_B(\zeta,y)= P_{B0}, \quad P_C(\zeta,y) =-
P_{B0}
$$
at $\zeta >0$. It means that the electromagnetic fields are
localized in waveguides without spreading to neighbor waveguides.
This kind of diffractionless propagation has been discussed and
observed in \cite{Mukherjee:15,Mukherjee:Spracklen:15a}.

By using the orthogonality condition
$$\int_{-\pi}^{\pi}e^{iy(n-m)}dy= 2\pi \delta_{n m},$$
the amplitudes $A_n(\zeta)$, $B_n(\zeta)$ and $C_n(\zeta)$ can be
determined from (\ref{eq:romb:Pro:sim:PA}),
(\ref{eq:romb:Pro:sim:PB}) and (\ref{eq:romb:Pro:sim:PC}):
$$
2\pi A_n(\zeta)=\int_{-\pi}^{\pi}P_A(\zeta, y)e^{-iyn}dy,\quad 2\pi
B_n(\zeta)=\int_{-\pi}^{\pi}P_B(\zeta, y)e^{-iyn}dy,\quad 2\pi
C_n(\zeta)=\int_{-\pi}^{\pi}P_C(\zeta, y)e^{-iyn}dy.
$$

\section{Some particular solution of the base equations}

\noindent Let us consider the particular case of the boundary
conditions for problem under consideration. The simplest case is
$$
A_n(0) = A_0\delta_{n 0}, \quad B_n(0) = B_0\delta_{n 0}, \quad
C_n(0) = C_0\delta_{n 0}.
$$
This conditions are correlated to situation where the radiation is
initially input only to waveguides of one elementary sell in array.
Thus,
$$
P_{A0} = A_0, \quad P_{B0} = B_0, \quad P_{C0} = C_0.
$$
By using the expressions (\ref{eq:romb:Pro:sim:PA}),
(\ref{eq:romb:Pro:sim:PB}) è (\ref{eq:romb:Pro:sim:PC}), the
amplitudes $A_n(\zeta)$, $B_n(\zeta)$ and $C_n(\zeta)$ can be
represented by following equations
\begin{eqnarray}
&& 2\pi A_n(\zeta)= A_0\int_{-\pi}^{\pi}\cos\Omega\zeta e^{-iyn}dy
-i\frac{R_0}{\sqrt{2}}  \int_{-\pi}^{\pi}\sin\Omega\zeta e^{-iyn-iy/2}dy, \label{eq:romb:sim:An} \\
&& 2\pi B_n(\zeta)= \frac{1}{2}S_0 \int_{-\pi} e^{-iyn}dy +
\frac{1}{2}R_0 \int_{-\pi}^{\pi}\cos\Omega\zeta e^{-iyn}dy
-\nonumber \\
&&\qquad \qquad  -\frac{iA_0}{\sqrt{2}} \int_{-\pi}^{\pi}
\sin\Omega\zeta e^{-iyn-iy/2}dy,  \label{eq:romb:sim:Bn}\\
&& 2\pi C_n(\zeta)= -\frac{1}{2}S_0 \int_{-\pi}e^{-iyn}dy +
\frac{1}{2}R_0 \int_{-\pi}^{\pi}\cos\Omega\zeta e^{-iyn}dy
-\nonumber \\
&&\qquad \qquad  -\frac{iA_0}{\sqrt{2}} \int_{-\pi}^{\pi}
\sin\Omega\zeta e^{-iyn-iy/2}dy, \label{eq:romb:sim:Cn}
\end{eqnarray}
Here the constants $S_0=B_0 -C_0$ and $R_0=B_0 +C_0$ are introduced.
Expressions (\ref{eq:romb:sim:An})--(\ref{eq:romb:sim:Cn})
demonstrates the principal role of the following integrals
\begin{eqnarray}
  I_n^{(1)} &=& \int_{-\pi}^{\pi}\cos\Omega\zeta e^{-iyn}dy = \int_{-\pi}^{\pi}\cos\left[2\sqrt{2}\zeta
  \cos(y/2)\right] e^{-iyn}dy,  \nonumber \\
  I_n^{(2)} &=& \int_{-\pi}^{\pi}\sin\Omega\zeta e^{-iyn-iy/2}dy =
  \int_{-\pi}^{\pi}\sin\left[2\sqrt{2}\zeta
  \cos(y/2)\right] e^{-iyn-iy/2}dy. \nonumber
\end{eqnarray}

To determine theses integrals the Anger's formula can be used
\cite{Bateman:Erdelyi:v2}. In the particular case this formula takes
the following form
\begin{eqnarray}
  &&\cos(z\cos\varphi) = J_0(z) +2\sum_{k=1}^{\infty}(-1)^kJ_{2k}(z)\cos(2k\varphi), \label{eq:romb:sim:Anger:7}\\
  &&\sin(z\cos\varphi) = 2\sum_{k=1}^{\infty}J_{2k-1}(z)\cos[(2k-1)\varphi]. \label{eq:romb:sim:Anger:8}
\end{eqnarray}
In this formulas substitutions $\varphi =y/2$ and $z=
2\sqrt{2}\zeta= \eta$ must be done.

Integral $I_n^{(1)}$ can be determined by using the
(\ref{eq:romb:sim:Anger:7}). Thus,
$$
I_n^{(1)}=\int_{-\pi}^{\pi}J_0(\eta)e^{-iyn}dy
+2\sum_{k=1}^{\infty}(-1)^kJ_{2k}(\eta)\int_{-\pi}^{\pi}\cos(ky)e^{-iyn}dy
=$$ $$ =J_0(\eta)\delta_{n 0}+\sum_{k=1}^{\infty}(-1)^kJ_{2k}(\eta)
\int_{-\pi}^{\pi}\left(e^{iy(k-n)}+ e^{-iy(k+n)}\right)dy = $$
$$ =2\pi J_0(\eta)\delta_{n 0}+2\pi\sum_{k=1}^{\infty}(-1)^kJ_{2k}(\eta)
\left(\delta_{k n}+ \delta_{-k n}\right).
$$
As only positive integer numbers we take into attention, the second
term in brackets is equal to zero. Thus,
\begin{equation}\label{eq:romb:sim:In1}
    I_0^{(1)}= 2\pi J_0(\eta), \quad I_n^{(1)}= 2\pi(-1)^nJ_{2n}(\eta), \qquad n\geq 1.
\end{equation}
As $J_{k}(z)=(-1)^kJ_k(z)$ is held, the expression
(\ref{eq:romb:sim:In1}) will be correct at negative integer
subindexes $n$.

Integral $I_n^{(2)}$ can be determined by using the
(\ref{eq:romb:sim:Anger:8}).
$$
I_n^{(2)}= -2\sum_{k=1}^{\infty}(-1)^kJ_{2k-1}(\eta)
\int_{-\pi}^{\pi} \cos[(2k-1)y/2]e^{-iy(n+1/2)}dy = $$
$$ = -\sum_{k=1}^{\infty}(-1)^kJ_{2k-1}(\eta)
\int_{-\pi}^{\pi}\left(e^{iy(k-1/2)}+
e^{-iy(k-1/2)}\right)e^{-iy(n+1/2)}dy = $$
$$= -\sum_{k=1}^{\infty}(-1)^kJ_{2k-1}(\eta)
\int_{-\pi}^{\pi}\left(e^{iy(k-n-1)}+ e^{-iy(k+n)}\right)dy =
$$ $$=-\sum_{k=1}^{\infty}(-1)^kJ_{2k-1}(\eta)2\pi\left(\delta_{k ~n+1}+
\delta_{k~-n}\right).
$$
The second term in brackets results in zero contribution. Thus,
\begin{equation}\label{eq:romb:sim:In2}
    I_n^{(2)}= -2\pi(-1)^{n+1}J_{2n+1}(\eta) = 2\pi(-1)^{n}J_{2n+1}(\eta), \qquad n\geq 0.
\end{equation}
With taking into account the expressions (\ref{eq:romb:sim:In1}) and
(\ref{eq:romb:sim:In2}) the amplitudes $A_n(\zeta)$, $B_n(\zeta)$
and $C_n(\zeta)$ can be written as
\begin{eqnarray}
&& A_n(\zeta)= A_0(-1)^nJ_{2n}(\eta) -\frac{iR_0}{\sqrt{2}}(-1)^nJ_{2n+1}(\eta), \label{eq:romb:sim:An:2} \\
&& B_n(\zeta)= \frac{1}{2}S_0 \delta_{n 0} + \frac{1}{2}R_0(-1)^n
J_{2n}(\eta) -\frac{iA_0}{\sqrt{2}}(-1)^nJ_{2n+1}(\eta), \label{eq:romb:sim:Bn:2}\\
&& C_n(\zeta)= -\frac{1}{2}S_0 \delta_{n 0} + \frac{1}{2}R_0(-1)^n
J_{2n}(\eta) -\frac{iA_0}{\sqrt{2}}(-1)^nJ_{2n+1}(\eta).
\label{eq:romb:sim:Cn:2}
\end{eqnarray}

In the case of boundary conditions
$$
           A_n(0) = A_0\delta_{n 0}, \quad B_n(0) = C_n(0) = 0,
$$
the expressions (\ref{eq:romb:sim:An:2})--(\ref{eq:romb:sim:Cn:2})
result in
\begin{eqnarray}
&& A_n(\zeta)= A_0(-1)^nJ_{2n}(\eta), \label{eq:romb:sim:An:3} \\
&& B_n(\zeta)= -\frac{iA_0}{\sqrt{2}}(-1)^nJ_{2n+1}(\eta), \label{eq:romb:sim:Bn:3}\\
&& C_n(\zeta)= -\frac{iA_0}{\sqrt{2}}(-1)^nJ_{2n+1}(\eta).
\label{eq:romb:sim:Cn:3}
\end{eqnarray}
The expressions (\ref{eq:romb:sim:An:3})--(\ref{eq:romb:sim:Cn:3})
describe the discrete diffraction (i.e., the electromagnetic
radiation  spreading along array).

In the case of boundary conditions $$ A_n(0) = 0, \quad B_n(0) = -
C_n(0) =B_0\delta_{n 0},
$$ we have $R_0=0$, but $S_0=2B_0$. The distribution of the electromagnetic fields in waveguide array
is
\begin{equation}\label{eq:romb:sim:ABCn:4}
   A_n(\zeta)= 0,\quad B_n(\zeta)= B_0 \delta_{n 0}, \quad C_n(\zeta)= -B_0 \delta_{n 0}.
\end{equation}
In this case the discrete diffraction is absent. It corresponds for
the flat-band. However, if the radiation will input into one of the
waveguide of the central part of array, i.e., to use the boundary
condition $$A_n(0) = A_0\delta_{n 0}, \quad B_n(0) = - C_n(0)
=B_0\delta_{n 0}, $$ than the distribution of the amplitudes
$A_n(\zeta)$, $B_n(\zeta)$ and $C_n(\zeta)$ takes the following form
\begin{eqnarray}
&& A_n(\zeta)= A_0(-1)^nJ_{2n}(\eta), \label{eq:romb:sim:An:5} \\
&& B_n(\zeta)= B_0 \delta_{n 0}-\frac{iA_0}{\sqrt{2}}(-1)^nJ_{2n+1}(\eta),  \label{eq:romb:sim:Bn:5}\\
&& C_n(\zeta)= -B_0 \delta_{n} -\frac{iA_0}{\sqrt{2}} (-1)^n
J_{2n+1}(\eta). \label{eq:romb:sim:Cn:5}
\end{eqnarray}
The discrete diffraction takes place in this case.

 \section{Conclusion}

\noindent The propagation of the electromagnetic continues wave in
the quasi-one-dimensional rhombic array of the waveguides is
investigated. The exact solution of the coupled mode equations
(\ref{eq:romb:ABCn:3}) is found by the use of generation function
method. In general case of the boundary conditions the discrete
diffraction is described by these solutions. However, the without
diffraction regime of the wave propagation at the particular
boundary condition exists. Taking into account the system of
equations (\ref{eq:romb:Pro:sim:PA})--(\ref{eq:romb:Pro:sim:PC}),
the interference of two discrete beams could be studied.

\section*{ Acknowledgement}

We are grateful to Prof. I. Gabitov and Dr. C. Bayun for
enlightening discussions. This investigation is funded by Russian
Science Foundation (project 14-22-00098).

\end{document}